\documentclass{jfm}
\usepackage{graphicx}
\usepackage{epstopdf, epsfig}
\usepackage[colorlinks,citecolor = black, linkcolor=black,hyperindex,CJKbookmarks]{hyperref}

\shorttitle{Rayleigh-Taylor instability in evaporating microdroplets}
\shortauthor{Y. Li, C. Diddens, T. Segers, H. Wijshoff, M. Versluis and D. Lohse}

\title{Rayleigh-Taylor instability by segregation in an evaporating multi-component microdroplet}

\author{Yaxing Li\aff{1},
  Christian Diddens\aff{1,2},
  Tim Segers\aff{1,2},
  Herman Wijshoff\aff{2,3},
  Michel Verluis\aff{1}
 \and Detlef Lohse\aff{1,4}
 \corresp{\email{d.lohse@utwente.nl}}
 }

\affiliation{\aff{1}Physics of Fluids group, Department of Science and Technology, Mesa+ Institute, Max Planck Center for Complex Fluid Dynamics and 
J. M. Burgers Centre for Fluid Dynamics, University of Twente, P.O. Box 217, 7500 AE Enschede, The Netherlands
\aff{2}Department of Mechanical Engineering, Eindhoven University of Technology, P.O. Box 513, 5600 MB Eindhoven, The Netherlands
\aff{3}Oc\'{e} Technologies B.V., P.O. Box 101, 5900 MA Venlo, The Netherlands
\aff{4}Max Planck Institute for Dynamics and Self-Organization, 37077 G\"ottingen, Germany
}

\begin{document}

\maketitle

\begin{abstract}
The evaporation of multi-component droplets is relevant to various applications but challenging to study due to the complex physicochemical dynamics. Recently, \cite{Li2018} reported evaporation-triggered segregation in 1,2-hexanediol-water binary droplets. In this present work, we added 0.5 wt\% silicone oil into the 1,2-hexanediol-water binary solution. This minute silicone oil concentration dramatically modifies the evaporation process as it triggers an early extraction of the 1,2-hexanediol from the mixture. Surprisingly, we observe that the segregation of 1,2-hexanediol forms plumes, rising up from the rim of the sessile droplet towards the apex during the droplet evaporation. By orientating the droplet upside down, i.e., by studying a pendant droplet, the absence of the plumes indicates that the flow structure is induced by buoyancy, which drives a Rayleigh-Taylor instability (i.e., driven by density differences \& gravitational acceleration). From $\mu$PIV measurement, we further prove that the segregation of the non-volatile component (1,2-hexanediol) hinders the evaporation near the contact line, which leads to a suppression of the Marangoni flow in this region. Hence, on long time scales, gravitational effects play the dominant role in the flow structure, rather than Marangoni flows. We compare the measurement of the evaporation rate with the diffusion model of~\cite{popov2005}, coupled with Raoult's law and the activity coefficient. This comparison indeed confirms that the silicone-oil-triggered segregation of the non-volatile 1,2-hexanediol significantly delays the evaporation. With an extended diffusion model, in which the influence of the segregation has been implemented, the evaporation can be well described.
\end{abstract}


\section{Introduction}

Evaporation of sessile droplets has attracted a lot of attention over the past decades due to its ubiquitousness and huge relevance for various applications, such as inkjet printing~\citep{park2006}, surface patterning~\citep{kuang2014}, microfabrication~\citep{kong2014}, among others. In particular, the pioneering work of \cite{deegan1997}, unveiling the mystery of the so-called coffee-stain effect, has inspired many scientific studies on evaporating droplets over the past twenty years.  

While the evaporation of single-component droplets is relatively well understood~\citep{hu2002,popov2005,ristenpart2007,cazabat2010,gelderblom2011,marin2011,lohse2015rmp,chong2020}, multi-component droplets show far more complex dynamics during the drying process. This is  due to the complicated coupling of the mutual interactions between species~\citep{Brenn2007,chu2016,diddens2017evaporating} and the resulting flow structures~\citep{kim2016controlled,Karpitschka2017,marin2019}. Essentially, the selective evaporation of each component is the reason underlying the complexity: the preferred evaporation of one component as compared to the other(s) can result in inhomogeneous liquid distributions. \cite{christy2011} first reported the sequential flow transitions in an evaporating ethanol-water binary droplet, which showed an evaporation-induced Marangoni instability in the early life stage. Many following studies~\citep{bennacer_sefiane_2014,zhong2016,diddens2017evaporating} show that the solutal Marangoni stress driven by the surface tension gradient dominates the flow structure in evaporating multicomponent microdroplets. Very recently, \cite{edwards2018} and \cite{Li2019} also found gravity-driven flows in different binary microdroplet systems, which are triggered by the density gradients from the selective evaporation. This is the first evidence that buoyancy-driven Rayleigh convection can overcome Marangoni flow in controlling the flow structure in such evaporating liquid-mixture droplets with Bond number Bo $\ll$ 1.

For a specific category of multicomponent systems with a metastable phase regime, the phenomena are even more intriguing and complex. \cite{Tan2016,Tan2017a,Tan2019NC} systematically studied a ternary ``Ouzo" system, which involves not only complex flow behaviours, but remarkably, multiple phase transitions, i.e., oil microdroplet nucleation and phase separation. Additionally, in a dissolution system, the interaction between host liquid and droplet liquids can also lead to segregation of the components inside the droplet~\citep{dietrich2017,Tan2019JFM}. Recently, \cite{Li2018} reported an unexpected segregation triggered by selective evaporation within a miscible 1,2-hexanediol-water binary droplet, in which 1,2-hexanediol is almost non-volatile compared to water. The insufficient replenishment of water from the droplet interior towards the contact line by the weak convection inside the droplet causes the local accumulation of 1,2-hexanediol in the contact line region, which eventually leads to segregation~\citep{Kim2018,karpitschka2018}.
 
In the current work, we added a small amount (0.5 wt\%) of silicone oil into the 1,2-hexanediol -water binary solution, which forms oil-water microemulsions in the mixture system~\citep{alany2000}, aiming to utilize silicone oil to trigger the extraction of 1,2-hexanediol. Surprisingly, we observed the plumes of separated 1,2-hexanediol arising along the droplet surface originating from the rim, which resemble those shapes immersing in a Rayleigh-Taylor instability~\citep{Rayleigh1882,Taylor1950}. To understand the mechanism of the plume formation and the evaporation behaviour of this multicomponent droplet,  
we studied the drying system experimentally and theoretically. 

The paper is organized as follows: In \S 2, we introduce the employed experimental methods. In \S 3, the experimental results and our interpretations thereof are presented. We then apply multicomponent-diffusion models to the experimentally analyzed cases (section 4). The paper ends with a summary and an outlook to the future work (section 5).

\section{Experimental methods}\label{sec:Exp_method}
\subsection{Solution and substrate}

The droplet system we used consisted of Milli-Q water (Reference A+, Merck Millipore, 25$^\circ$C), 1,2-hexanediol (Sigma-Aldrich; $\geqslant$ 98$\%$) and silicone oil (Sigma-Aldrich, viscosity 1000 cSt). First, 10 wt\% of 1,2-hexanediol aqueous solution was prepared and then 0.5 wt\% of silicone oil was added. For this we mix 30~mg silicone oil with 5970~mg 1,2-hexanediol aqueous solution in a glass container and sonicate it for 10~minutes. We performed evaporation experiments on a hydrophobized glass slide coated by Octadecyltrichlorosilane (OTS, $>$ 90 \%, Sigma-Aldrich)~\citep{Peng2014}. Before usage, the substrates were cleaned by 15-min sonication in 99.8\% ethanol and 5 min in Milli-Q water sequentially and subsequently dried with compressed N$_2$ flow for 30 sec. The droplet in each experiment was deposited by a glass syringe with a full metal needle (Hamilton, 10 $\mu$L, Model 701 NWG SYR, Cemented NDL).

\subsection{Confocal microscopy}

Confocal microscopy was employed to visualize the distribution of water
and 1,2-hexanediol within the mixture droplet. The observations were
carried out by using an inverted Nikon A1 confocal laser scanning microscope system (Nikon
Corporation, Tokyo, Japan) with a 10$\times$ dry objective (Nikon, Plan Fluor 10 ×/0.30, OFN25,
DIC, L/N1). The droplet was
labeled with two different dyes, i.e., Nile Red and Dextran. Nile Red is a lipophilic dye
which dissolves only in 1,2-hexanediol and was excited by a laser at a wave length of 561 nm, while Dextran preferentially dissolves in water and it was excited by a laser at a wave length of 488 nm simultaneously. Three-dimensional (3D) images were obtained by reconstruction from a series of consecutive Z-stack images scanned in the direction from the substrate to the top of the droplet. The scan started as soon as the droplet was deposited on the glass substrate. Operating in Galvano mode,
the scan rate for the 2D images was 1~fps, while each Z-stack scan loop for 3D images took
approximately 30 s to complete. This timescale is much smaller than that of evaporation
such that the variation of the flow pattern within the droplet during the scans was negligible. 

\subsection{Micro particle image velocimetry}

For flow visualization, we performed micro Particle Image Velocimetry ($\mu$PIV) by adding
fluorescent particles [Fluoro-Max; Red Fluorescent Polymer Microspheres: Ex/Em 530
nm/607 nm; Diameter: 0.52 $\mu$m] into the working fluids at a concentration of 2 $\times$ 10$^{-2}$ vol\%. The $\mu$PIV measurements were implemented on the same confocal microscope with a 10$\times$ dry objective (Nikon, Plan Fluor 10$\times$/0.30, OFN25,
DIC, L/N1). The particles were excited by a laser at a wave length of 561 nm and
the fluorescent signals were captured at a frame rate of 25 fps. The
droplet was illuminated from the bottom and the fluorescent signal was also captured by the objective from the bottom.

\subsection{Geometrical measurement}

The evaporation process was recorded by a CCD camera [MQ013MG-E2, XiQ] coupled to a microscope [12X Ultrazoom, NAVITAR], which was illuminated by LED light [MWWHL4 Warm White Mounted LED, THORLABS] from the opposite side of the droplet. In the experiments, droplets evaporated into air under stable laboratory conditions. The relative humidity $RH$ and the ambient temperature $T$ were monitored in each measurement, which were 45 $\pm$ 3\% and 22 $\pm$ 1$^\circ$C, respectively. 
 
\section{Rayleigh-Taylor instability arising from segregation}\label{sec:RT}

\subsection{Experimental observations and interpretations}

\begin{figure}
\centering
  \includegraphics[width=1\textwidth]{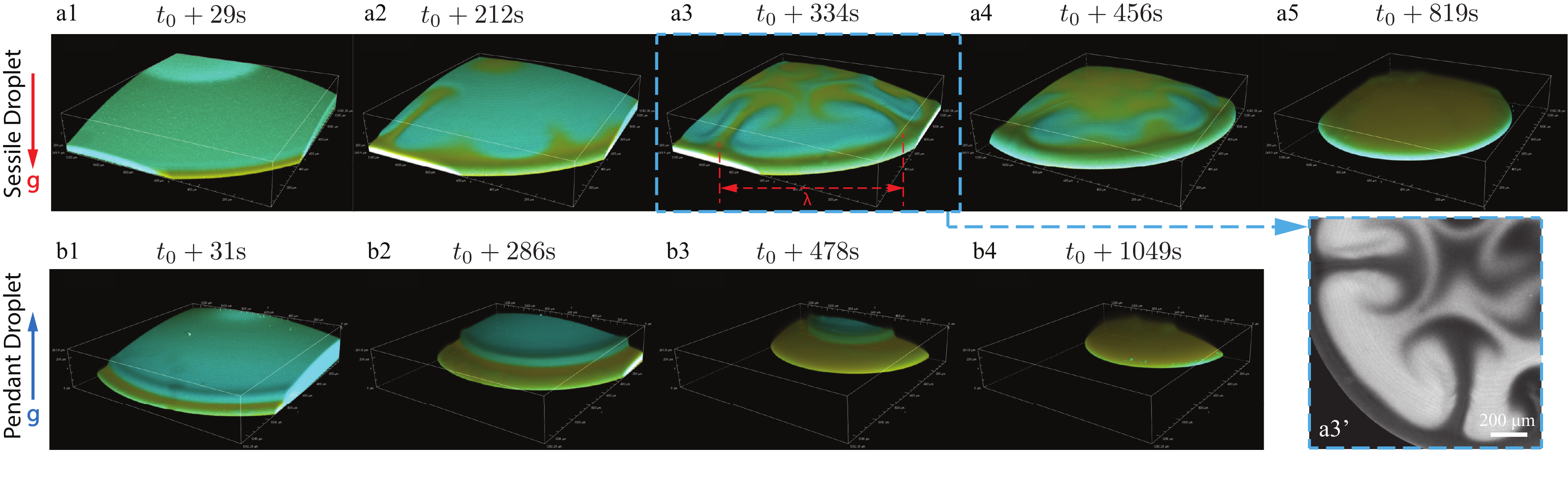}
  \caption{Confocal images of evaporation behaviours for both sessile (a1-a5) and pendant (b1-b4) droplets in a semi-side view taken at different time instants. The confocal microscope scanned the rectangular box with the volume 1225 $\mu$m $\times$  1280 $\mu$m $\times$ 250 $\mu$m. (a1),(b1) For both droplets, when the evaporation began, the 1,2-hexanediol separated at the contact line and formed a ring-like pattern. (a2-a3) In the sessile droplet, through the growth of the segregation, the separated 1,2-hexanediol rose up with plumes. (a3') The top view of the droplet at $t_0 + 334$s. The figure is transformed into binary image to increase the contrast of colours. (a4) The plumes merged with each other at the apex of the droplet. (a5) Eventually, 1,2-hexanediol fully covered the surface and stopped the evaporation. (b2 - b3) In the pendant droplet, the segregation of 1,2-hexanediol expanded with the ring-like shape. (b4) Once the separated 1,2-hexanediol occupied the entire surface area, the evaporation stopped.}
  \label{fig:RT}
\end{figure}

Evaporation processes of silicone-oil-seeded 1,2-hexanediol-water droplets with opposite orientations are displayed in Fig.~\ref{fig:RT}. The upper (a1-a5) and lower (b1-b4) row show the evolution of a sessile droplet and a pendant droplet, respectively. In the beginning of the recording (approximately 30 sec after droplet deposition), the segregation of 1,2-hexanediol already appeared for both droplets [Fig.~\ref{fig:RT}(a1,b1)], as revealed by the yellow colour.  

\begin{figure}
\centering
  \includegraphics[width=1\textwidth]{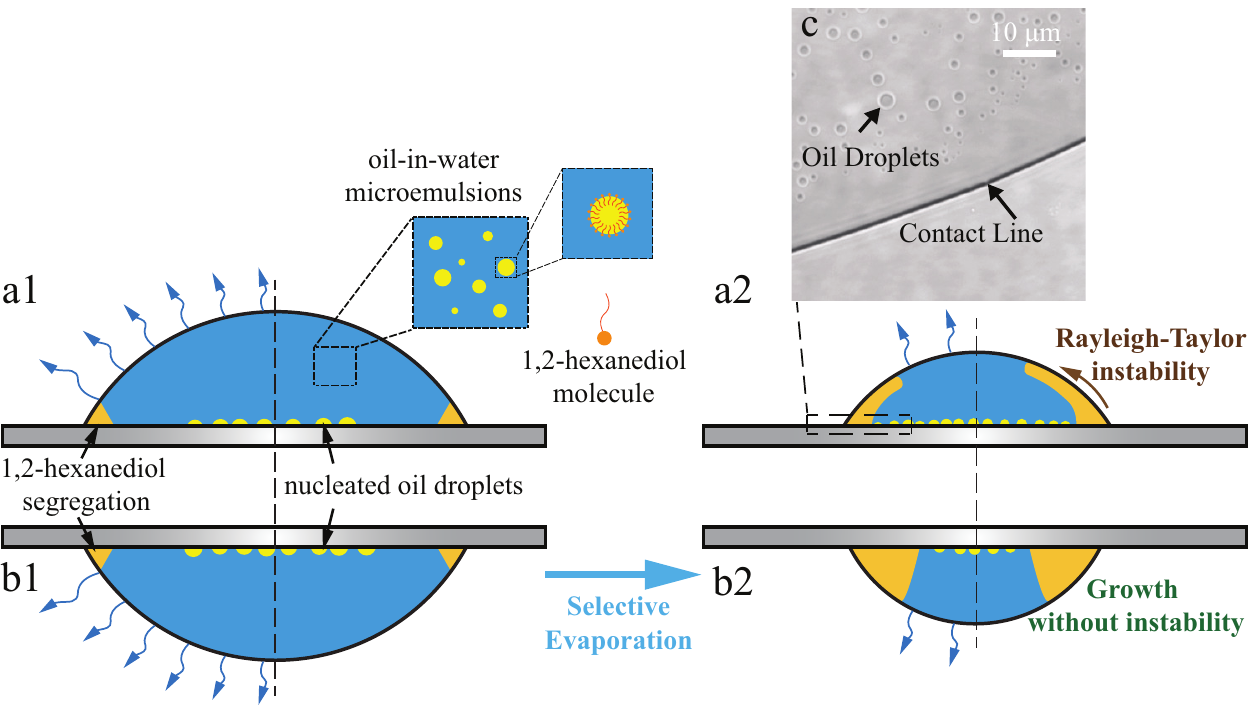}
  \caption{Schematics of the silicone-oil-seeded binary droplets with opposite orientations. (a1) Within the bulk of the droplet, there are oil-water microemulsions. Because of the preferential evaporation of water near the contact line, 1,2-hexanediol is extracted by silicone oil and starts separating in this region. The non-volatile 1,2-hexanediol segregation shields the evaporation of water at the rim. (a2) The weak surface tension gradient cannot lead to a strong Marangoni flow on the surface. Instead, in the long term, buoyancy drives the arising plumes (Rayleigh-Taylor instability). (b1) When we orientate the droplet upside down, a similar segregation of 1,2-hexanediol occurs near the contact line. (b2) However, the segregation rim continuously grows due to the selective evaporation of water but no plumes appear due to the inverted direction of gravity. (c) The nucleated oil droplets on the substrate (bottom optical view).}
  \label{fgr:sketch}
\end{figure}

For the sessile droplet in Fig.~\ref{fig:RT}(a2), in the segregation process, plumes formed, rising from the contact line towards the apex of the droplet. In Fig.~\ref{fig:RT}(a3), the shape of the plumes resembles those shapes immersing in a Rayleigh-Taylor instability~\citep{sharp1984}. The plumes later coalesce with each other, eventually fully covering the whole surface of the droplet and thereby ceasing the evaporation process. However, for the pendant droplet, the separated ring only monotonically grows without any plume formation. This observation clearly demonstrates that the flow structure on the droplet interface is controlled by gravity. The mechanism of the formation of the buoyancy-driven flow structure is interpreted along the schematics in Fig.~\ref{fgr:sketch}. In equilibrium, the silicone oil forms oil-in-water microemulsions owing to the existence of 1,2-hexanediol as a surfactant~\citep{alany2000}. After the droplet being deposited on the substrate, some oil droplets nucleate on the solid surface, as was also observed experimentally in Fig.~\ref{fgr:sketch}(c). The reason is that the silicone oil has a much lower interfacial energy with OTS glass than the 1,2-hexanediol-water mixture. This can be seen by comparing the equilibrium contact angle of a pure silicone oil droplet on the OTS glass with that of the 1,2-hexanediol-water mixture droplet. The former is only 10$^\circ$, the latter is 40$^\circ$. Early on in the evaporation process, the oil droplets in the oil-water emulsions in the bulk of the droplet aggregate due to the depletion of water. These oil droplets together with those which already nucleated on the substrate trigger the extraction of 1,2-hexanediol from the aqueous solution such that the 1,2-hexanediol phase separates from the water phase.
 
For a sessile droplet drying on a flat substrate with a contact angle smaller than 90$^\circ$, the evaporative flux is maximal at the contact line~\citep{deegan1997}. Hence, the extraction starts from the edge of the droplet due to the fast evaporation of water in that region, leading to an 1,2-hexanediol ring that hinders the further evaporation from the contact line region. The non-volatile ring impedes the building up of the concentration gradient, which results in a suppression of the surface tension gradient, which therefore cannot play any dominant role in controlling the flow. Instead, because of the lower density of 1,2-hexanediol as compared to the mixture, the separated phase of 1,2-hexanediol at the bottom of the droplet rises up to the apex, driven by buoyancy.

\subsection{Evidence of suppression of Marangoni flow from $\mu$PIV measurements}

\begin{figure}
\centering
  \includegraphics[width=1\textwidth]{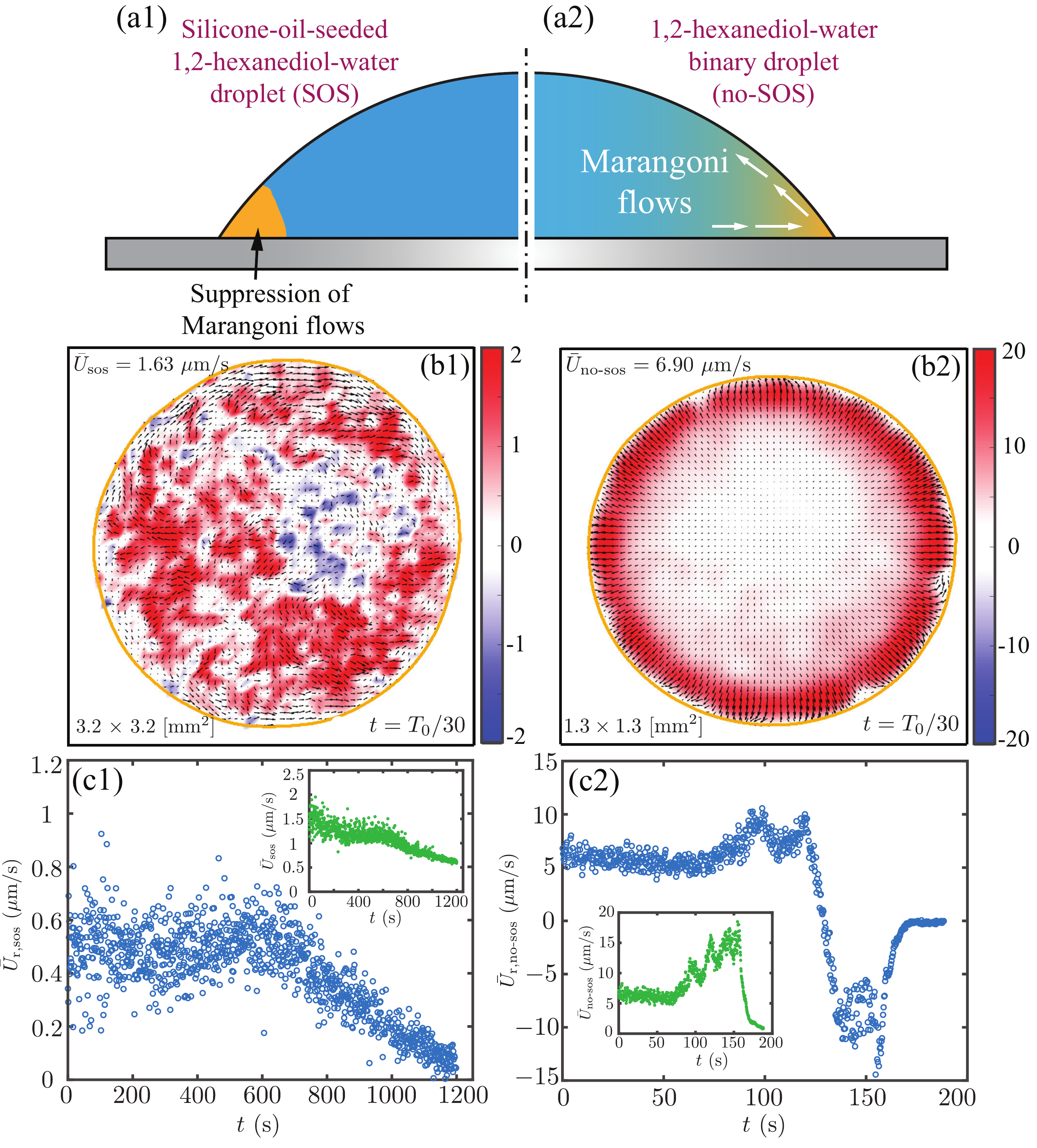}
  \caption{$\mu$PIV measurements of the velocity fields of a silicone-oil-seeded 1,2-hexanediol-water (SOS) droplet and a 1,2-hexanediol-water (no-SOS) binary droplet~\citep{Li2018}. (a1),(a2) Schematics of both droplets at the early life stage. (b1),(b2) $\mu$PIV snapshot of the velocity field in the focal plane near the substrate at the beginning of the evaporation process. The arrows display the local velocity and the radial velocity is colour coded. (b1) The map shows that there is no visible coherent radial flow. Note that the color scale bar for $U_\mathrm{r,sos}$ ranges from~-2~to~+2~$\mu$m/s, indicating a weak Marangoni flow in a SOS droplet. (b2) The liquid flows radially towards the edge of the no-SOS droplet from the interior. The radial flow is most intense ($\sim 20\ \mu$m/s) near the contact line, implying a strong Marangoni flow there. The color scale bar for $U_\mathrm{r,no-sos}$ covers a 10 times larger range as that for $U_\mathrm{r,sos}$. (c1),(c2) The evolution of the radial velocity $U_\mathrm{r}$ in the focal plane near the substrate for both droplets. The inlets show the evolution of absolute mean flow velocity $U$.}
  \label{fig:velocity}
\end{figure}

To prove our interpretation, we performed $\mu$PIV measurements to characterize the flow field within the evaporating silicone-oil-seeded (``SOS") 1,2-hexanediol-water droplet. Fig.~\ref{fig:velocity}(a1) schematically illustrates that the segregation of 1,2-hexanediol suppresses the Marangoni flow in the contact line region. Fig.~\ref{fig:velocity}(b1) displays a snapshot of the velocity field in the focal plane near the substrate at an early stage of the evaporation process, $t = T_0/30$ ($T_0$ is the droplet's lifetime). The velocity map shows chaotic and very weak flow motions. Fig.~\ref{fig:velocity}(c1) shows the evolution of the mean radial velocity and the absolute mean velocity (inlet plot). The mean radial velocity $\bar{U}_\mathrm{r,sos}$ is less than 1 $\mu$m/s, and the absolute mean velocity $\bar{U}_\mathrm{sos}$ in the early stage is around $1\ \mu$m/s. For comparison, we also measured the flow field for a 1,2-hexanediol-water binary droplet (without the silicone-oil seeding), which reveals the absence of density-driven flow~\citep{Li2018}. In that case the velocity map in Fig.~\ref{fig:velocity}(b2) shows much more intense outward radial flows close to the contact line than that in the case with silicone-oil seeding. During the early lifetime, the mean radial flow velocity $\bar{U}_\mathrm{r,no-sos}$ [Fig.~\ref{fig:velocity}(c2)] is more than 5 $\mu$m/s, which is one order of magnitude higher than that in the former case. The measured velocities for both cases imply Reynolds numbers Re$_\mathrm{sos}$ = $\rho R_d\bar{U}_\mathrm{sos}/\mu \sim 10^{-4}$ and Re$_\mathrm{no-sos}$ = $\rho R_d\bar{U}_\mathrm{no-sos}/\mu \sim 10^{-3}$, where $\rho \approx 10^3\ \textrm{kg/m}^3$, $R_d \approx 10^{-3}\ \textrm{m}$, and $\mu \approx 10$ mPa$\cdot$s are the liquid density, droplet radius, and liquid viscosity, respectively. We estimate the Marangoni time scale in the two cases by using the mean velocity of the radial flow: $t_\mathrm{Ma,sos} \sim R_d/\bar{U}_\mathrm{r,sos} \approx 10^{-3} \textrm{m}/10^{-6} \textrm{m/s} = 10^3$ s and $t_\mathrm{Ma,no-sos} \sim R_d/\bar{U}_\mathrm{r,no-sos} \approx 10^{-3} \textrm{m}/10^{-5} \textrm{m/s} = 10^2$ s. By looking at the rising time of the plumes from Fig.~\ref{fig:RT}, we obtain the Rayleigh time scale of the RT instability $t_\mathrm{Ra} \sim 10^2$ s (estimated time of the plumes rising up from the rim to the apex of the droplet). In the silicone-oil-seeding case, the Rayleigh time scale is much smaller than the Marangoni time scale: $t_\mathrm{Ra}/t_\mathrm{Ma,sos} \ll 1$, which indicates that the buoyancy flow is dominant. In the non-silicone-oil-seeding case, the two time scales are comparable: $t_\mathrm{Ra}/t_\mathrm{Ma,no-sos} \approx 1$, which substantiates that gravity-driven flow is balanced by Marangoni flow, thereby playing no controlling role. We argue that in the silicone-oil-seeded 1,2-hexanediol-water droplet, the instantaneous segregation hinders the evaporation near the contact line, which suppresses the most intensive Marangoni flow in that region, leading to a weak flow motion in the whole droplet. Therefore, on a relatively long time scale, the buoyancy force due to the density difference dominates the flow, causing Rayleigh-Taylor instability. 

We estimate the most unstable wavelength of the RT instability in our system $\lambda_m \approx 4\pi [(\nu^2/(g_s\mathrm{At})]^{1/3} \approx 10^{3}\ \mu$m~\citep{olson2009}. In this expression, $g_s = g\textrm{sin}(\theta)$ is the net acceleration imposed on the interface of the droplet, where $g \approx 9.8\ \textrm{m/s}^2$ is the gravitational acceleration and $\theta \approx 35^\circ$ is the droplet contact angle [see Fig.~\ref{fgr:theta_radius}(d1,d2)]; $\mathrm{At}$ is the Atwood number given by $\mathrm{At} = (\rho_m - \rho_H)/(\rho_m + \rho_H) \approx 2.3\times10^{-2}$, where $\rho_m = 997$ kg/m$^{3}$ and $\rho_H = 952$ kg/m$^{3}$ are the density of the mixture and of pure 1,2-hexanediol, respectively~\citep{romero2007} and $\nu = (\mu_m + \mu_H)/(\rho_m + \rho_H)$ is the averaged kinematic viscosity, where $\mu_m \approx 2$ mPas and $\mu_H \approx 80$ mPas are the dynamic viscosities of the mixture and of pure 1,2-hexanediol, respectively~\citep{Jarosiewicz2004}. Here $\mathrm{At} \ll 1$, the low density liquid which is 1,2-hexanediol, moves into the heavy fluid in the upper layer~\citep{sharp1984}. The estimated wavelength $\lambda_m$ is comparable to the spatial distance between two plumes $\lambda \approx 10^3\ \mu$m in Fig.~\ref{fig:RT}(a3), supporting our interpretation of the plumes as RT instability. 

\subsection{Evaporation-triggered extraction of 1,2-hexanediol by seeding oils}

\begin{figure}
\centering
  \includegraphics[width=1\textwidth]{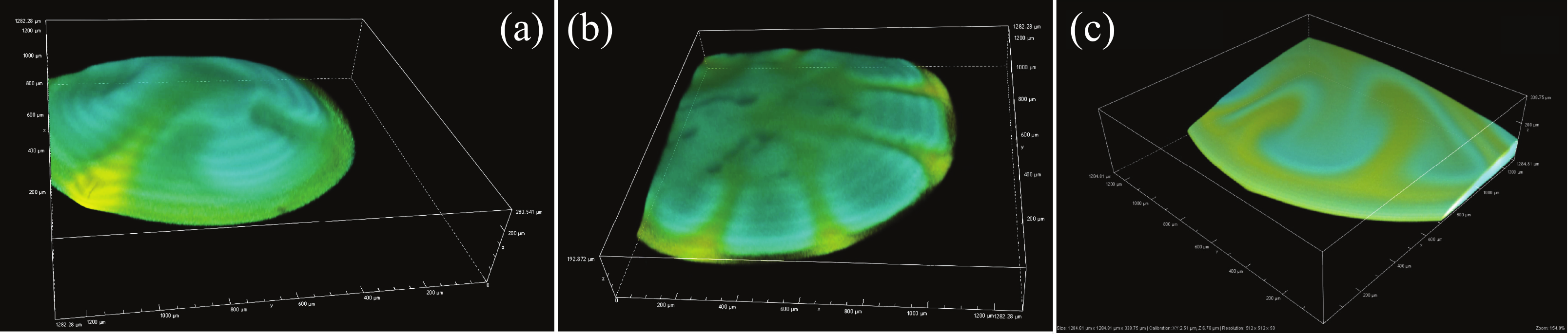}
  \caption{Confocal images of segregation patterns for droplets seeded with different oils, namely (a) silicone oil with 12500 cSt viscosity, (b) silicone oil with 100 cSt viscosity, and (c) 1,2-hexanediol-water binary droplet deposited by a plastic syringe and a disposable needle. The confocal microscope scan covered a rectangular box with the section area of 1225~$\mu$m~$\times$~1280~$\mu$m.}
  \label{fgr:diff_oil}
\end{figure}

To evaluate the applicability of different oils for the extraction effect, we seeded the 1,2-hexanediol-water binary solution with several kinds of oils at the concentration of 0.5 wt\%, namely silicone oils with viscosities of 12500 cSt and 100 cSt. As shown in Fig.~\ref{fgr:diff_oil}(a,b), the 1,2-hexanediol-water droplets seeded with the three different oils all show similar plumes rising up from the rim of the sessile droplet towards the apex during the evaporation process. The consistency clearly demonstrates that the evaporation can trigger the early extraction of 1,2-hexanediol by the oil-water emulsions in these solutions, leading to the segregation of 1,2-hexanediol and the resulting flow structures.

We also tested a series of concentration ratios of the seeding oil: 0.5 wt\%, 0.1 wt\%, 0.05 wt\%, and 0.025 wt\%. Our observation shows that the effect holds for the concentration of silicone oil down to 0.025 wt\%. The robustness of the extraction effect even at tiny silicone oil concentrations obviously rises the question on the oil-contamination control in such liquid systems. This issue had been addressed before by~\cite{berkelaar2014} in a study on nanobubble nucleation. They found that the polydimethylsiloxane (PDMS) coating in a plastic syringe and a disposable needle can contaminate the solution to form nanobubble-like objects on the substrate, which in fact were silicone oil nanodroplets. In their study, the concentration ratio between PDMS (Sylgard 184, Dow Corning) and water was 0.1 mL/0.4 L, which is $\approx 0.025$ wt\%.

To see whether we can trigger similar contamination effects also here, in a test experiment, we used the same plastic syringe (5 mL, Discardit, BD) and disposable needle (Microlance, BD) to deposit a 1,2-hexanediol-water \textit{binary} droplet on the substrate. Indeed, similar RT-like patterns as in our other experiments with silicone oil also occur in such a droplet during the evaporation process [as shown in Fig.~\ref{fgr:diff_oil}(c)], which confirms that even minute PDMS contamination can cause the early extraction of 1,2-hexanediol during the evaporation process.

\subsection{Reversed segregation by evaporation on a lubricated film}

In the previous sections, we experimentally demonstrated that gravitational effects dominate the flow structure in the droplet system, due to the suppression of Marangoni flow by the instantaneous segregation of 1,2-hexanediol close to the contact line. We argued that the early phase separation is caused by the coupling of the extraction effect by the seeding oil and the maximal evaporation rate at the contact line. In order to validate this argument, we suppress the local evaporative flux at the contact line by introducing a non-volatile wetting ridge, which can be achieved by letting the droplet evaporate on a lubricated surface~\citep{Schellenberger2015,gao2019}.

We performed the experiment in which the silicone-oil-seeded droplet evaporates on a lubricated surface of silicone oil (Sigma-Aldrich, viscosity 1000 cSt). The lubricated surface was made by spin coating of silicone oil on a solid glass substrate (Gerhard Menzel GmbH, 76 $\times$ 26 mm), with a typical thickness of 18 $\pm\ 1\ \mu$m~\citep{hack2018}. The equilibrium contact angle $\theta$ of the droplet on this lubricated surface is 38$^\circ$ [see Fig.~\ref{fgr:theta_radius}(d1,d2)], which is close to the contact angles in the first two cases. In the experiment, we still only dyed the 1,2-hexanediol and water with the same method mentioned in \S2. Thus the silicone oil film is not visible in the confocal movie. 

\begin{figure}
\centering
  \includegraphics[width=1\textwidth]{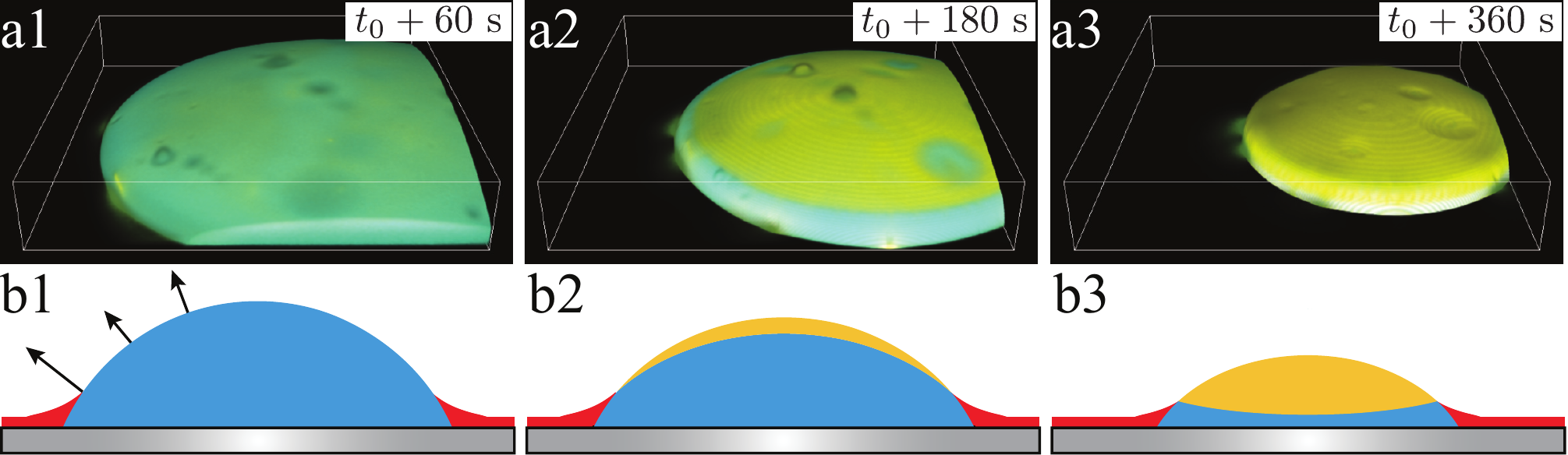}
  \caption{The dynamic behaviour of a silicone-oil-seeded 1,2-hexanediol/water droplet evaporating on a silicone oil thin film. (a1-a3) Confocal microscopy images for a scanned box with volume 1225 $\mu$m $\times$  1280 $\mu$m $\times$ 250 $\mu$m. (a1) In the beginning, the droplet is homogeneously mixed. (a2) The segregation of 1,2-hexanediol started appearing on the upper surface of the droplet instead of contact line area. (a3) The evaporation ceased when the droplet surface was shielded by 1,2-hexanediol. (b1-b3) Schematic of the evaporation process. The red colour represents the silicone oil thin film. It forms a meniscus at the contact line.}
  \label{fgr:on_thin_film}
\end{figure}

One wonders whether there is an intercalated film in between the droplet and the substrate. To find out, we performed interferometry~\citep{daniel2017}. The measurements do not show any interference patterns, which suggests that there is no stable intercalated film. Within a second, the droplet rewets the substrate by rupturing the thin film after the deposition. We estimate the spreading coefficient $S = \gamma_\mathrm{da} - \gamma_\mathrm{do} - \gamma_\mathrm{oa} \approx 24~\mathrm{mN/m} - 20~\mathrm{mN/m} - 21~\mathrm{mN/m} < 0$, where $\gamma$ is the interfacial tension, and the subscripts a, o, and d indicate the air and oil phases, and the droplet, respectively. The result $S~<~0$ is consistent with no oil-engulfment covering on the surface of the droplet. In the beginning of the evaporation [Fig.~\ref{fgr:on_thin_film}(a1)], the droplet is homogeneously mixed, which reveals the green colour. Later on, as shown in Fig.~\ref{fgr:on_thin_film}(a2), the segregation of 1,2-hexanediol (yellow colour) starts appearing at the upper part of the droplet rather than from the edge. Also note that in this third case the segregation behavior is different from the first two cases, as it appears more homogeneously and slowly. The reason is that the segregation in this third case is triggered by the nucleated oil droplets in the bulk instead of those on the substrate, which have a much higher number density in the early phase of the nucleation process and which play the essential role to trigger the early segregation. In Fig.~\ref{fgr:on_thin_film}(a3), the evaporation ceases with the shielding of separated 1,2-hexanediol. This observation shows a different route of segregation, which indicates a faster evaporation rate from the upper surface than from the contact line. 

The explanation is that the thin film forms a wetting ridge covering on the contact line region, as illustrated in Fig.~\ref{fgr:on_thin_film}(b1), which hinders the evaporative flux from there~\citep{gao2019}. Consequently, only the water molecules on the upper surface of the droplet evaporate to the surrounding air, which leads to a high concentration of 1,2-hexanediol at the top of the droplet. Then the highly concentrated 1,2-hexanediol on the upper surface nucleates and segregates from the mixture [Fig.~\ref{fgr:on_thin_film}(b2)]. In the end of the evaporation, there is still water entrapped by the shielding thanks to the segregated 1,2-hexanediol and the silicone oil meniscus, as illustrated in Fig.~\ref{fgr:on_thin_film}(b3). 

\begin{figure}
\centering
  \includegraphics[width=1\textwidth]{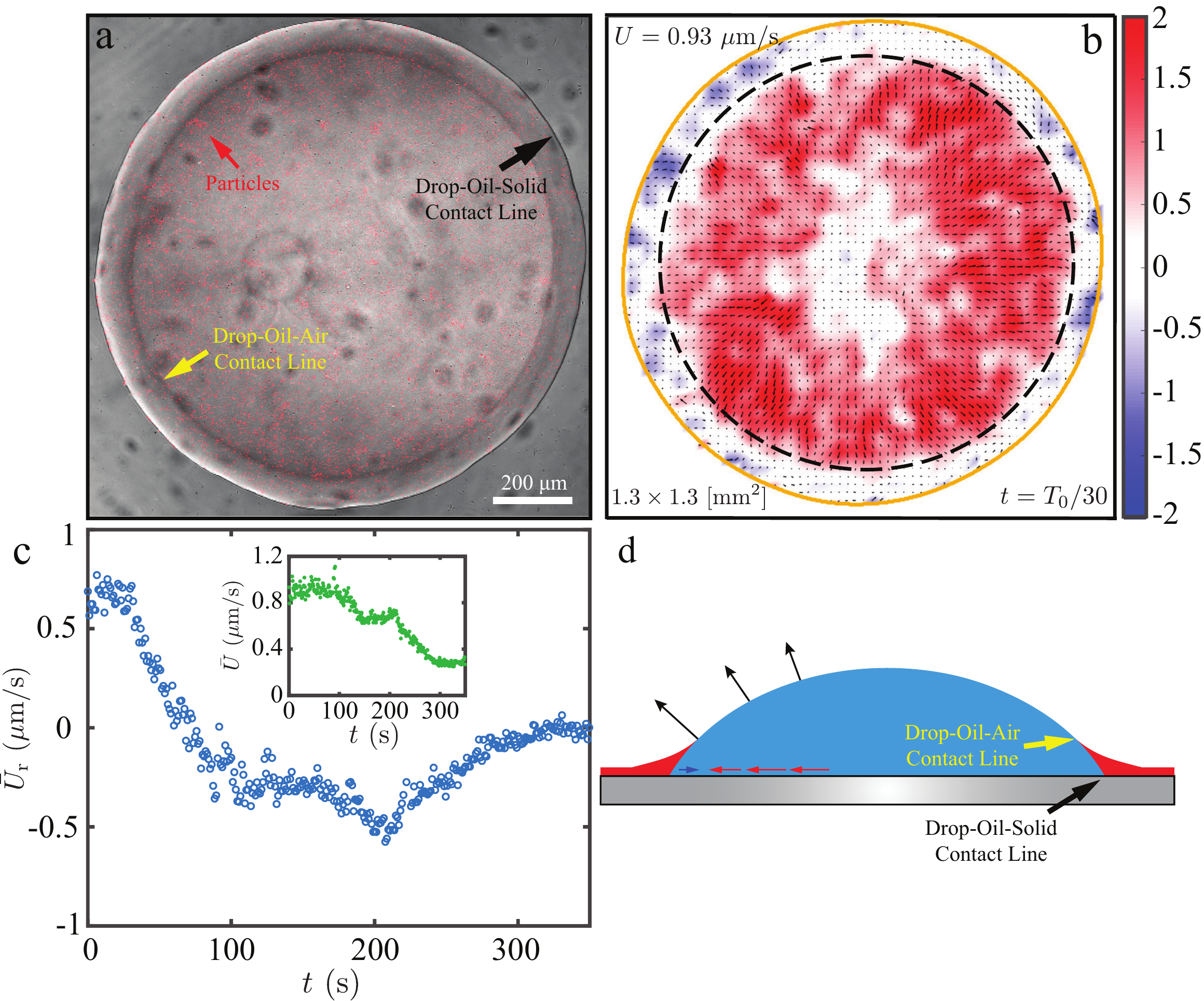}
  \caption{$\mu$PIV measurement of a silicone-oil-seeded 1,2-hexanediol/water droplet evaporating on a silicone oil thin film. (a) $\mu$PIV image focusing on the bottom of the droplet ($\approx 10\ \mu$m above the substrate). Note that the two circular rings indicate the drop-oil-solid contact line (marked by the black arrow) and the drop-oil-air contact line (marked by the yellow arrow). (b) Flow field of $\mu$PIV at $t = T_0/30$, where $T_0$ is the droplet's lifetime. The black dashed line indicates the horizontal position of the drop-oil-air contact line. (c) The evolution of the mean radial velocity $\bar{U}_\mathrm{r}$ and the mean absolute velocity $\bar{U}$. (d) Schematics of a silicone-oil-seeded 1,2-hexanediol/water droplet evaporating on a silicone oil thin film.}
  \label{fgr:PIV_thinfilm}
\end{figure}

To further suggest the argument that the wetting ridge suppresses the local evaporative flux at the contact line, we also employed $\mu$PIV measurement for the silicone-oil-seeded 1,2-hexnaediol/water droplet evaporating on a silicone oil thin film, as shown in Fig.~\ref{fgr:PIV_thinfilm}. The flow field was measured at the focal plane $\approx$~10~$\mu$m above the substrate. Fig.~\ref{fgr:PIV_thinfilm}(a) displays the bottom-view image at the beginning of the evaporation process. Two circular rings can be observed. As indicated by the black and the yellow arrows, the outer ring and inner ring represent the drop-oil-solid contact line and the drop-oil-air contact line, respectively. Fig.~\ref{fgr:PIV_thinfilm}(b) shows the typical velocity field of the flow structure. The radial velocity $U_\mathrm{r}$ is colour-coded. One can see that the radial flow is comparable to the radial velocity in the first case [Fig.~\ref{fig:velocity}(b1)], which is much weaker as compared to the Marangoni flow in the binary droplet [Fig~\ref{fig:velocity}(b2)]. Moreover, the outward radial flow from the interior ceases at the horizontal position of drop-oil-air contact line. Near the contact line of the droplet, the radial flow even reverses inwardly [revealed by blue colour in Fig.~\ref{fgr:PIV_thinfilm}(b)]. The flow reversal reflects that the liquid in the contact line region can only flow inwardly to maintain the decreasing contact angle and the pinning contact line at this early stage of the droplet's lifetime [see Fig.~\ref{fgr:theta_radius}(d1, e1)]. This is a direct evidence that the evaporative flux near the contact line is suppressed by the oil wetting ridge. As shown in Fig.~\ref{fgr:PIV_thinfilm}(c), the mean radial flow decreases and then changes the direction due to the shrinking of the contact area [Fig.~\ref{fgr:theta_radius}(d1, e1)].

\section{Evaporation process and its quantitative understanding}

A thorough insight in the evaporation process requires a quantitative understanding. Although the evaporation of multicomponent droplet is far more complicated than single-component droplet, the essence of the system is still limited by the diffusion of vapor molecules of each component to the surrounding air~\citep{Langmuir1918,diddens2017evaporating}. In this session, we first introduce a diffusion model to describe the evaporative rate for multicomponent droplets. Then we show the geometrical measurement for the three different cases: a. a silicone-oil-seeded 1,2-hexanediol-water sessile droplet; b. a silicone-oil-seeded 1,2-hexanediol-water pendant droplet; and c. a silicone-oil-seeded 1,2-hexanediol-water droplet on a lubricated surface. Finally, we compare a modified multicomponent-diffusion model with the influence of segregation considered with our measurement and discuss the results.

\subsection{Multicomponent-diffusion model}
\cite{popov2005} proposed an analytical description for the diffusion-controlled evaporation of a sessile droplet with one single component, which was later experimentally confirmed by~\cite{gelderblom2011,Sobac2011}. For the evaporation of multicomponent droplets, we use the method put forward by~\cite{Brenn2007}, considering the total evaporation rate of the mixture droplet as the sum of the evaporation rates of each individual component. In our droplet mixture system, only the diffusive flux of water contributes to the total evaporation rate. However, non-volatile components in the system also affect the vapor-liquid equilibrium:  the existence of 1,2-hexanediol and silicone oil change the saturated concentration of water vapor at the liquid-air interface. Raoult's law is used to calculate the difference: $c_{w,s} = X_w c_{w,s}^{0}$, where $X_w$ is the liquid mole fraction of the water component at the interface and $c_{w,s}^{0} = 2.08 \times 10^{-2}$ kg/m$^3$ is the saturated vapor concentration of the pure water at room temperature. However, Raoult's law only relies on an ideal solution and thus ignores any interaction between the compositions. To overcome this limitation, a so-called activity $a_i$ for each component is introduced to describe this interaction~\citep{chu2016}, $a_i = \psi_i X_i$, where $\psi_i$ is the activity coefficient. In our case, the mole fraction of silicone oil is negligible ($<$~0.1~\%), we only consider the interaction between water and 1,2-hexanediol. Therefore the saturated concentration of water vapor at the interface becomes $c_{w,s} = a_w c_{w,s}^{0}$. By using Raoult's law together with the water activity $a_{w}$~\citep{marcolli2005} to modify the one-liquid model~\citep{popov2005}, we obtain a theoretical model to express the evaporation rate for the water in our system:

\begin{equation}
\centering
\frac{\textrm{d}M}{\textrm{d}t}=-\pi DR(a_w c_{w,s}^{0}-c_{w,\infty})f(\theta),
\label{eq:modified_popov}
\end{equation}
with

\begin{equation}
\centering
f(\theta)=\frac{\textrm{sin}(\theta)}{1+\textrm{cos}(\theta)}+4\int_{0}^{\infty} \frac{1+\textrm{cosh}(2\theta\varepsilon)}{\textrm{sinh}(2\pi\varepsilon)}\textrm{tanh}[(\pi-\theta)\varepsilon] \textrm{d}\varepsilon,
\end{equation}
where $D = 24.6 \times 10^{-6}\ \mathrm{m}^2/$s is the diffusion coefficient of water vapour in air at room temperature, $R$ and $\theta$ are the footprint radius and contact angle of the droplet, respectively.  Besides controlling the evaporation rate, the model also yields the terminal state of the evaporation, which is when the saturated water vapor concentration equals the environmental concentration, $c_{w,s} = c_{w,\infty}$. Theoretically, the evaporation ceases when the active mole fraction of water equals to the relative humidity of the surrounding air, $a_w = \psi_w X_w= RH$. From the relative humidity $RH$ measured in each experiment, we can analytically calculate the ``theoretical final volume" $V_f$ (as shown in appendix~\ref{appA}) of each measured droplets as:

\begin{equation}
V_f = \left(\frac{M_w}{M_H}\frac{RH}{\psi_w-RH}+\frac{\rho_w}{\rho_H}\right)\left(\frac{1-C}{C}+\frac{\rho_w}{\rho_H}\right)^{-1}V_0,
\label{eq:final}
\end{equation}
in which $M_H$ and $M_w$ are the molecular mass of 1,2-hexanediol and that of water, $\rho_H$ and $\rho_w$ are their liquid densities at room temperature, and $C$ is the initial mass concentration of 1,2-hexanediol in each measurement. We rescale the measured droplet volume and time, by introducing non-dimensional volume $\hat{V} = (V-V_f)/(V_0-V_f)$ and time $\hat{t} = t/\tau_c$~\citep{gelderblom2011}, in order to compare the different sets of experimental data. $V$ is the droplet volume measured in every time interval, $V_0$ is the initial volume of each measurement and $V_f$ is the estimated final volume by Eq.~(\ref{eq:final}). $\tau_c$ is the characteristic timescale of droplet lifetime, $\tau_c = \rho_{l}R^2/(D\Delta c)$~\citep{gelderblom2011}, where $\rho_{l}$ is the density of the liquid and $\Delta c$ the water vapor concentration difference between the air-liquid interface and the surrounding air.

\subsection{Evaporation modes and volumetric measurement}
\begin{figure}
\centering
  \includegraphics[width=1\textwidth]{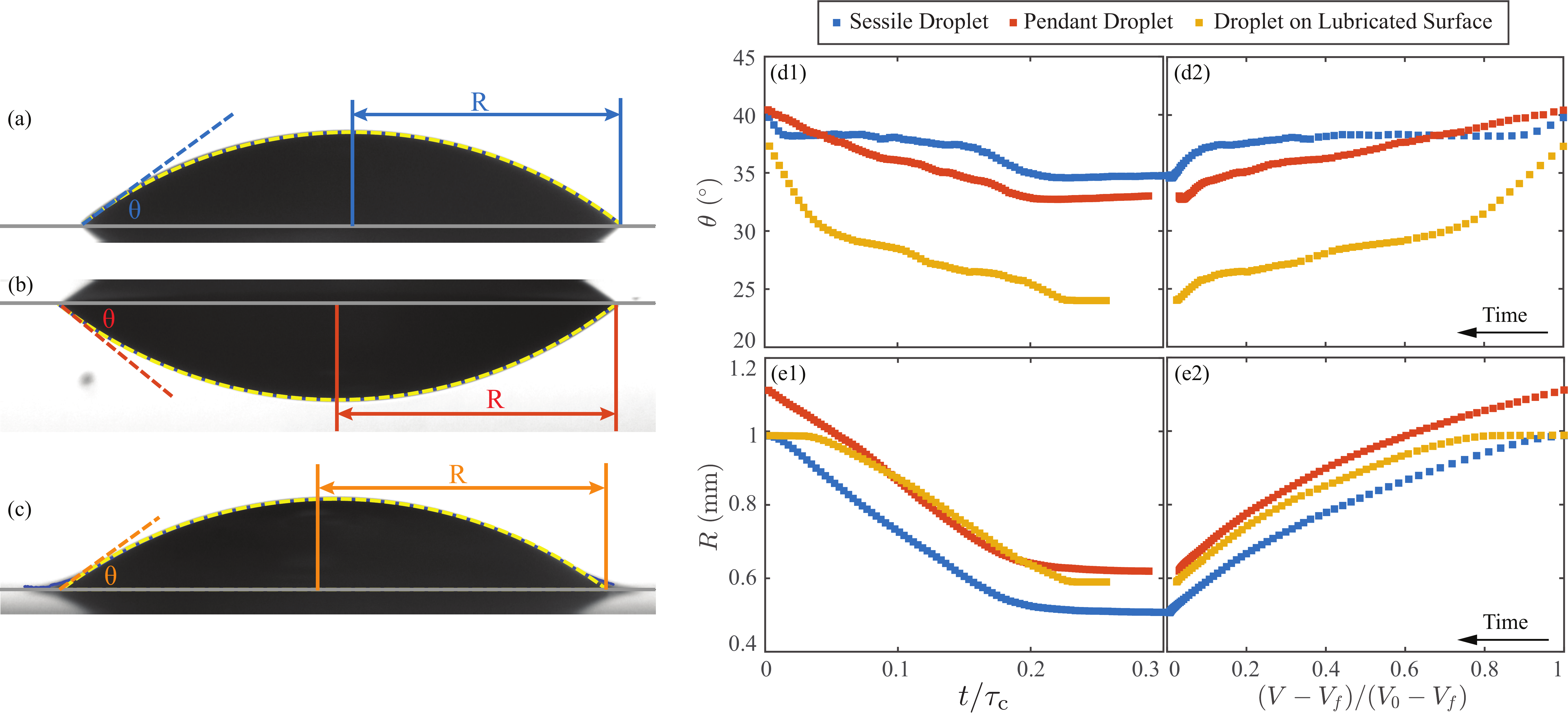}
  \caption{Morphology evolution of the evaporating droplets in three different scenarios: sessile, pendant and on the lubricated film. (a-c) Recorded images of the droplets in the three cases, with annotations of the geometrical parameters, i.e., contact angle $\theta$ and footprint radius $R$. Note that for the droplet on lubricated surface (c), we fit the large part of the surface with a spherical curve (see the yellow dashed line), and define the contact angle and footprint radius of the fitting shape as $\theta$ and $R$. (d-e) The contact angle $\theta$ and footprint radius $R$ as a function of scaled time $\hat{t} = t/\tau_c$ and scaled droplet volume $\hat{V} = (V - V_f)/(V_0 - V_f)$.}
  \label{fgr:theta_radius}
\end{figure}

\begin{figure}
\centering
  \includegraphics[width=1\textwidth]{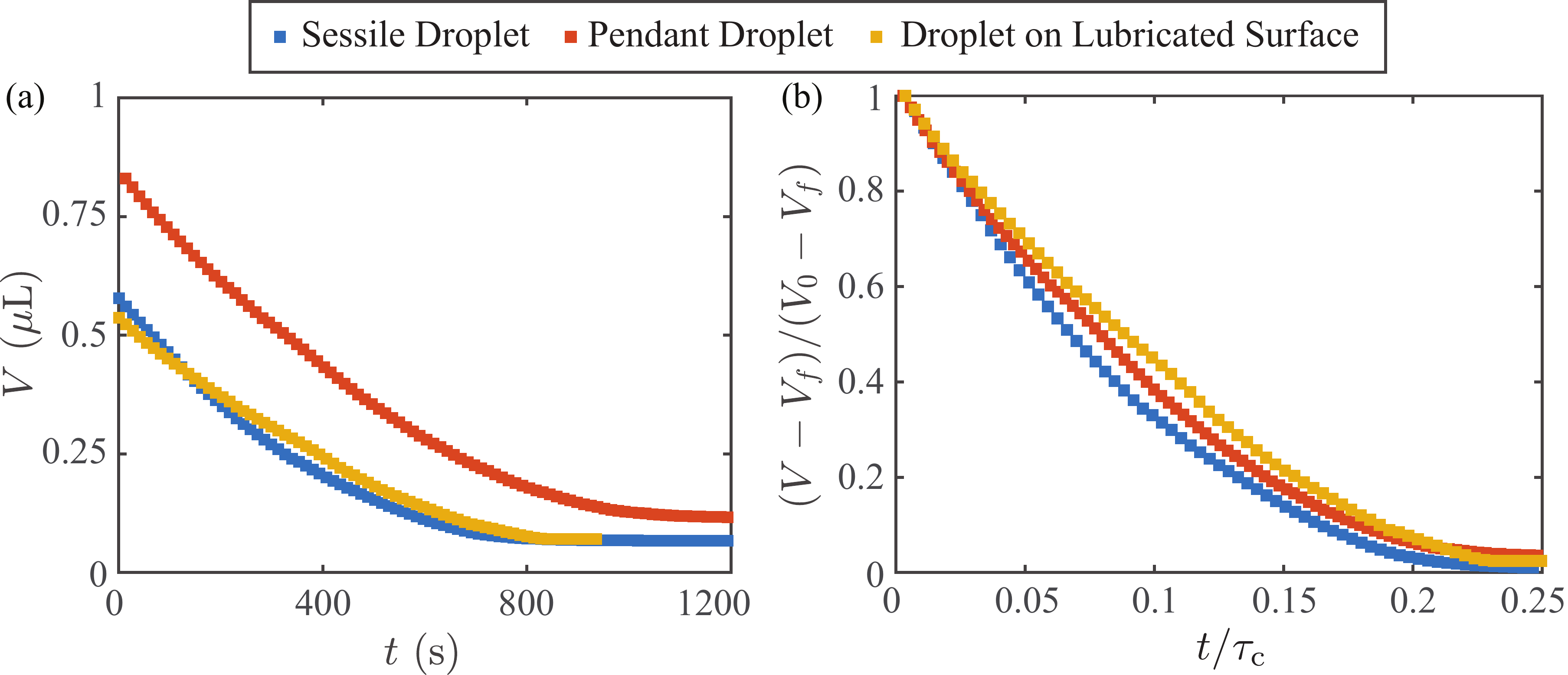}
  \caption{(a) Volumetric evolutions of a sessile droplet, a pendant droplet and a droplet on a lubricated surface. (b) Scaled droplet volume $\hat{V} = (V - V_f)/(V_0 - V_f)$ as a function of scaled time $\hat{t} = t/\tau_c$.}
  \label{fgr:volume}
\end{figure}

In Fig.~\ref{fgr:theta_radius} we show that the measured contact angle $\theta$ and the footprint radius $R$ of (a) a sessile droplet, (b) a pendant droplet and (c) a droplet on a lubricated surface. Figure.~\ref{fgr:theta_radius}(a-c) show a snapshot of each droplet. Note that due to the existence of the oil meniscus in the third case [Fig.~\ref{fgr:theta_radius}(c)], we define the contact angle and footprint radius by fitting the droplet profile with a spherical cap, shown as the yellow dashed line. Figs.~\ref{fgr:theta_radius}(d-e) show the evolution of the two parameters as a function of the scaled time $\hat{t} = t/\tau_c$ and volume $\hat{V} = (V - V_f)/(V_0 - V_f)$. We observe a sliding contact line during the evaporation for both sessile and pendant droplets. For the sessile droplet, the contact angle remains almost constant during most of the lifetime, but for the pendant droplet, the contact angle always decreases. For the droplet on the thin film, the evaporation follows the stick-slide mode~\citep{stauber2014,stauber2015,Nguyen2012}, i.e., it first evaporates with a pinning contact line, and then shrinks with decreasing contact angle. In Fig.~\ref{fgr:volume}(a), we show the volumetric evolutions for the three cases. The evaporation lifetime is normally affected by both the evaporation modes and environmental conditions. In Fig.~\ref{fgr:volume}(b) we compare the volumetric evolution for the three cases by  rescaling them with $\hat{t} = t/\tau_c$ and $\hat{V} = (V - V_f)/(V_0 - V_f)$. The deviation between the curves clearly implies that the evaporation kinetics is affected not only by the evaporation modes and environmental conditions, but also by the segregation patterns. Note that the first case with the Rayleigh-Taylor instability [blue squared dots in Fig.~\ref{fgr:volume}(b)] shows the fastest volumetric decrease among the three cases (compare the slopes of the curves). The percentages of entrapped water compared to the initial amount of water in the final residual of the three droplets were around 16\%, 30\%, and 27\%, respectively, estimated from the final volume of droplets. We argue that the fastest evaporation rate coinciding with the lowest water entrapment percentage in the first case is because of the Rayleigh-Taylor instability, causing a better mixing of the components than in the other two cases.

\subsection{Evaporative flux profile and modified diffusion model}

\begin{figure}
\centering
  \includegraphics[width=0.8\textwidth]{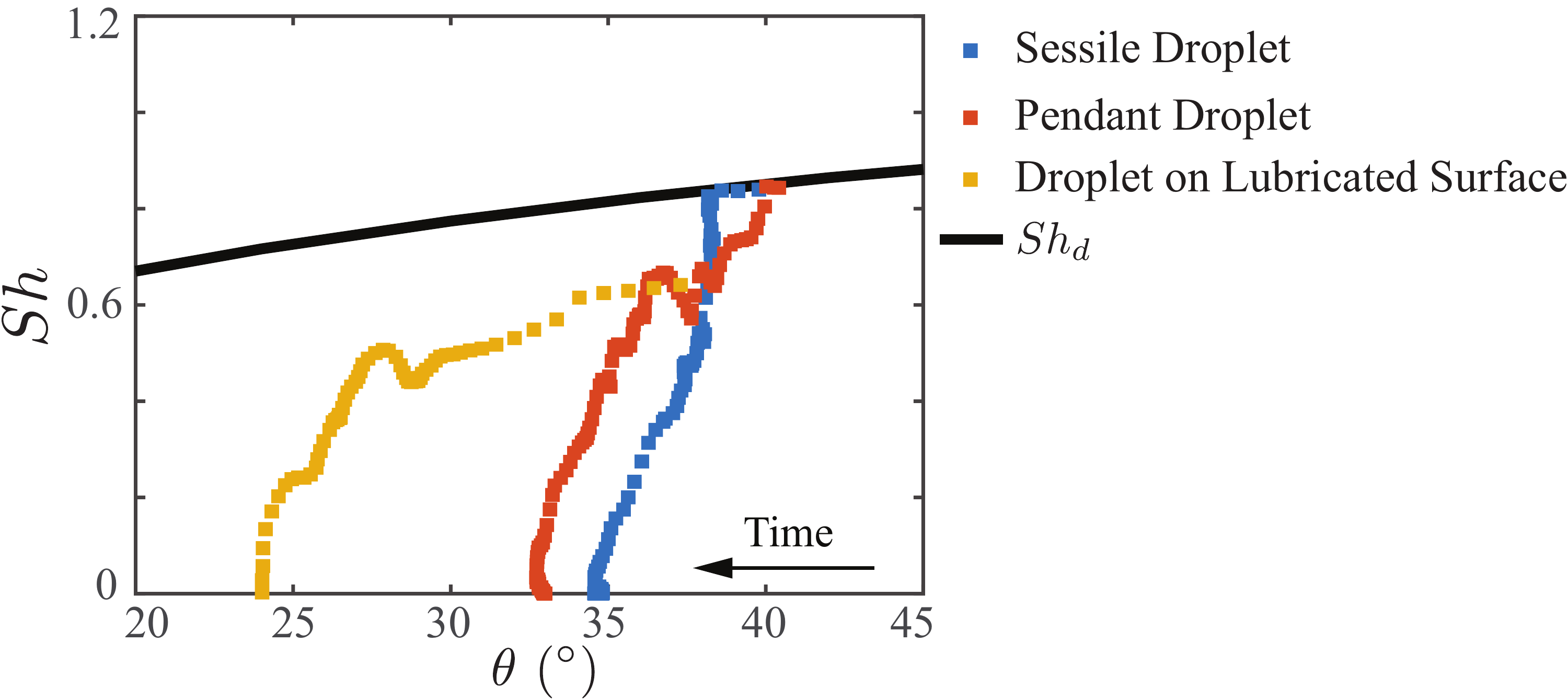}
  \caption{Sherwood number as a function of contact angle of a sessile droplet (blue), a pendant droplet (red), and a droplet on a lubricated surface (yellow). The black solid line represents the theoretical Sherwood number $Sh_d$, which is described by Eq.~(\ref{eq:shd}).}
  \label{fgr:dMdt}
\end{figure}

To further quantify the mass transfer process, we first use the Sherwood number $Sh$ derived by \cite{dietrich2016} in a diffusion-limited case, and extend it to mixture system by including Raoult's law and activity coefficient :
\begin{equation}
\centering
Sh = \frac{\langle\dot{M}\rangle_A R_{eq}}{D(a_w c_{w,s}^{0}-c_{w,\infty})},
\label{eq:sh}
\end{equation}
where $\langle\dot{M}\rangle_A$ is the actual mass loss rate (measured in experiments), averaged over the droplet surface area $A$, and $R_{eq} = (3V/(2\pi))^{1/3}$ is the equivalent radius. For sessile droplets, if the mass transfer occurs purely via diffusion, the Sherwood number is~\citep{dietrich2016} 
\begin{equation}
\centering
Sh_d = \frac{f(\theta)}{\textrm{sin}\theta (1 + \textrm{tan}^2\frac{\theta}{2}) (\frac{2}{2-3\textrm{cos}\theta + \textrm{cos}^3\theta})^{1/3}},
\label{eq:shd}
\end{equation}
which only depends on the droplet contact angle $\theta$. Note that the Sherwood number here scales mass transfer with respect to a diffusive spherical droplet. In Fig.~\ref{fgr:dMdt}, we plot the experimental data for the three cases by rescaling them based on Eq.~(\ref{eq:sh}) with an assumption that the mixture liquid components homogeneously distribute. The overestimation of the Sherwood number following from the single-component diffusive model (black curve) as compared to the experiments indicates that the assumption of homogeneous mixing does not hold for the three cases, which clearly is due to the segregation of 1,2-hexanediol. Specifically, for both sessile (blue squared dots) and pendant (red squared dots) droplets, the experimental data and theoretical model show good agreement at the beginning of the evaporation process and they deviate from each other later on. The reason for this deviation is that the distribution of compositions is no longer homogeneous when segregation of 1,2-hexanediol occurs, i.e., the assumption of an evenly mixed system overestimates the local concentration of water on the surface. For the droplet on a lubricated surface (yellow squared dots), the model overpredicts the evaporation rate for the whole process, even from the very beginning. This is due to the fact that the non-volatile oil meniscus hinders the evaporation of water near the contact line during the entire droplet lifetime.

\begin{figure}
\centering
  \includegraphics[width=1\textwidth]{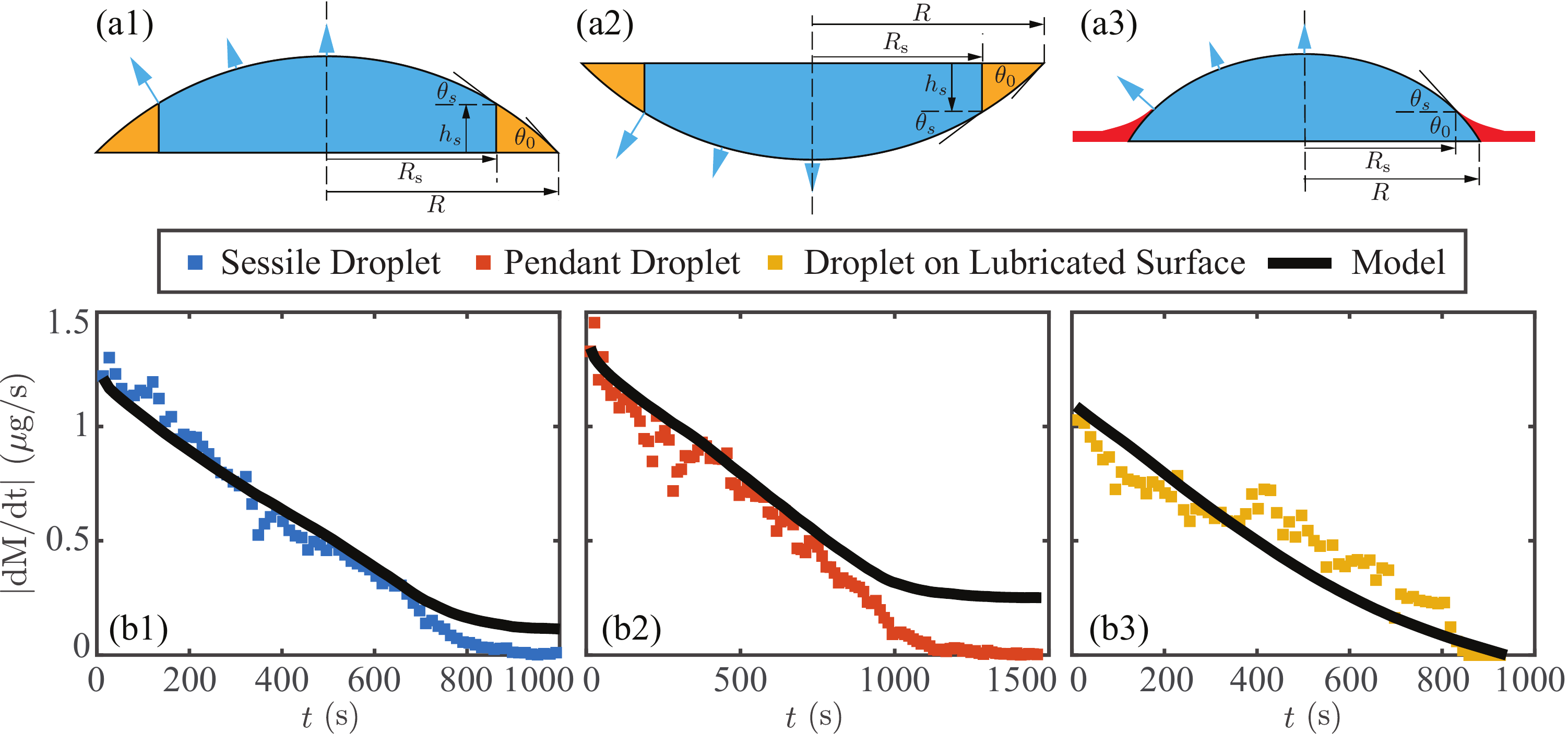}
  \caption{Schematic view of the evaporation of (a1) a sessile droplet, (a2) a pendant droplet, and (a3) a droplet on a lubricated surface. (b1-b3) Temporal evaporation rate of each droplet in (a1-a3), respectively, during the evaporation process. The black solid line represents the prediction of the theoretical model.}
  \label{fgr:evap_pro}
\end{figure}

To get a quantitative description of the evaporation rate in the mixture droplets, we further modify the mixture model [Eq.~(\ref{eq:modified_popov})] by considering the effect of segregation patterns on the evaporative flux profile of each droplet. For the first two cases, as sketched in Fig.~\ref{fgr:evap_pro}(a1,a2), the hindrance of the local evaporative flux originates from the non-volatile 1,2-hexanediol segregation. In the limit of small contact angle ($\theta < 40^\circ$), the local evaporative flux on the drop-air interface can be described as $j(r,\theta) \sim (R_s - r)^{-\frac{\pi-2\theta}{2\pi - 2\theta}}$, which is maximal at $r = R_s$~\citep{deegan1997,sobac2014}. $R_s$ is the distance between the center of the droplet and the front of the segregation. Hence we obtain the total evaporative flux of the droplet with segregation by replacing $R$ with $R_s$ in Eq.~(\ref{eq:modified_popov}):

\begin{equation}
\centering
\frac{\textrm{d}M}{\textrm{d}t} = \int^{R_s}_{0} j(r,\theta)\sqrt{1+(\partial_rh)^2}2\pi r dr = -\pi DR_s(a_w c_{w,s}^{0}-c_{w,\infty})f(\theta_s),
\label{eq:modified_popov2}
\end{equation}
where $\theta_s$ is the tangential angle at the front of the segregation [see Fig.~\ref{fgr:evap_pro}(a1)]. As the angle is small, $0 \leq \theta_s \leq 40^\circ$, we approximate $f(\theta_s) \approx 4/\pi$~\citep{hu2002,Sobac2011} in our calculation for convenience. This approximation was shown to be appropriate in colloidal suspension droplets with similar configurations~\citep{sobac2014}. In order to calculate $R_s$, the volume of the segregated liquid is required. From the confocal imaging, we observe an almost instantaneous segregation of 1,2-hexanediol. Hence we assume that the water mostly depletes in the segregation region [the yellow part in Fig.~\ref{fgr:evap_pro}(a1)] and remains nearly constant in the mixture region [the blue part in Fig.~\ref{fgr:evap_pro}(a1)]. The mass in the segregation region $\Delta M_s$ can be calculated by multiplying the volume of the region $\Delta V_s$ and the density of 1,2-hexanediol $\rho_H$:

\begin{equation}
\Delta M_s = \rho_H \Delta V_s = \rho_H \int^{R}_{R_s} 2\pi rh(r) dr
=\frac{\pi \theta}{4R}(R^2-R^2_s)^2,
\label{eq:separated_ring_1}
\end{equation}
where $h(r) = \frac{R^2-r^2}{2R}\theta$ is the local height calculated from the parabolic approximation at distance $r$ from the center of the droplet. The separated 1,2-hexanediol is caused by the depletion of water $\Delta M_w = \rho_w \Delta V$, where $\Delta V$ is the volume loss measured from the experiment and $\rho_w$ is the density of water. According to the initial mass concentration of 1,2-hexanediol $C$, we then have:

\begin{equation}
\Delta M_s = \frac{C}{1-C}\Delta M_w = \frac{C}{1-C}\rho_w\Delta V.
\label{eq:depleted_water}
\end{equation}

By combining Eq.~(\ref{eq:separated_ring_1}) and Eq.~(\ref{eq:depleted_water}), we obtain

 \begin{equation}
R_s = R\left(1-\sqrt{\Delta V\frac{C}{1-C}\frac{\rho_w}{\rho_H}\frac{4}{\pi \theta R^3}}\right)^{1/2}.
\label{eq:Rs}
\end{equation}

By substituting Eq.~(\ref{eq:Rs}) into Eq.~(\ref{eq:modified_popov2}), we can theoretically calculate the evaporation rate without any adjustable parameters. Note that we approximate $a_w(X_w)$ by taking the initial water mole fraction $X_w$ as a constant for the whole process. For the third case where the droplet evaporates on the thin film, the evaporative flux profile is changed by the hindrance of the oil wetting ridge. As shown in Fig.~\ref{fgr:evap_pro}(a3), the evaporative flux profile is identical to the first two cases: it is maximal at the drop-oil-air contact line. Therefore, the total evaporation flux rate can also be described by Eq~(\ref{eq:modified_popov2}). $R_s$ then is the horizontal distance from the center of the droplet to the triple-phase contact line. We can measure $R_s$ by determining the position of the triple-phase contact line from the bottom-view image, as shown in Fig.~\ref{fgr:PIV_thinfilm}(a). We use $a_w(X_w)$ with a temporal water mole fraction $X_w$ calculated from the volumetric measurement.

In Fig.~\ref{fgr:evap_pro}(b1-b3), we show the measured evaporation rate corresponding to each case and compare them with the theoretical prediction by Eq.~(\ref{eq:modified_popov2}). For both sessile and pendant droplets [Fig.~\ref{fgr:evap_pro}(b1-b2)], the experimental data and theoretical model show good agreement for a large part of the process and deviate from each other only in the late stage of the lifetime. The main reason for this deviation at the very end of the evaporation process is that the entrapped water by the segregation of 1,2-hexanediol does not homogeneously distribute in the droplet, i.e., the assumption of an evenly mixed system overestimates the local concentration of water on the surface. For the droplet on a lubricated surface, the model gives a good description of the evaporation rate.

Finally note that in the second half of the lifetime, the model underestimates the evaporation rate. This can be explained by the fact that near the triple-phase contact line, the wetting edge is not thick enough to completely hinder the evaporation of water. Water can still evaporate into air in that region by first diffusing through the thin layer of oil.

\section{Summary and outlook}

In this work, we experimentally studied the evaporation behaviour of a silicone-oil-seeded 1,2-hexanediol-water droplet. The observation shows an instantaneous segregation of 1,2-hexanediol in the sessile droplet  followed by the formation of plumes of the segregated fluid. By orientating the droplet upside down, the absence of the plumes indicates that the observed instability is indeed controlled by gravity, confirming the interpretation as Rayleigh-Taylor instability. We have shown that through hindering the strong evaporation near the contact line by segregation of non-volatile 1,2-hexanediol, the Marangoni effect can be significantly suppressed, and thus allowing buoyancy force to play a dominant role in the flow structure. Following the idea of suppressing evaporation locally by coverage with a non-volatile component, we can manipulate the segregation to start preferentially from the top rather than the edge of the droplet by letting the droplet evaporate on a lubricated surface.

Since the evaporation of droplets shows complex physicochemical behaviour, i.e., segregation of 1,2-hexanediol, it is difficult to obtain knowledge of the local concentration. We theoretically apply a scaling analysis on the experimental investigation of droplets in three cases, namely the sessile droplet, the pendant droplet and the droplet on a lubricated surface. It shows different evaporation modes of contact line behaviour in each case. By comparing the measurement of the evaporation rate with the multicomponent-diffusion model, we show that the segregation of the non-volatile component significantly delays the evaporation of water and, even leading to entrapment of water in the residue of the droplets.

To our best knowledge, this work is the first observation of the Rayleigh-Taylor instability in a microdroplet system triggered by evaporation. It is another example that the Rayleigh convection can overcome Marangoni effects to control the flow structure in a milli-sized droplet with Bond number $< 1$. As we have demonstrated that the mixing effect is highly influenced by the flow pattern, our finding can be crucial for many applications involving uniform surface coating and particle assembly.  We have also shown that such a surprising phenomenon can be triggered by various different seeding oils, even at very low concentrations. We think that it is important for the community to realize that during evaporation processes which involve several different components, each component even with minute amount may dramatically influence the overall behavior.

Moreover, droplets need not always evaporate on a solid surface: There are many applications involving droplets drying on lubricated surfaces. In particular, in the inkjet printing process, a primary layer is printed on surfaces prior to the deposition of ink drops, in order to destabilize the pigments to improve the printing quality~\citep{hack2018}. Our study of a silicone-oil-seeded mixture droplet evaporating on a lubricated surface provides an effective way to manipulate the segregation in such drying systems by utilizing the non-volatile meniscus to impede the evaporation from the edge.

Many questions still remain open. The difficulty to accurately predict the droplet lifetime originates from the lack of means to accurately monitor the local distribution of the liquid components. The complexity arises not only from the segregation of 1,2-hexanediol, but also the later plume formation. How to predict the onset of the Rayleigh-Taylor instability in this geometry of spherical cap? Does the unstable wave length between the plumes depend on the size or the contact angle of the droplets? To answer these questions, further studies on the parameter space are worthwhile. By considering the limitation of experimental measurement, a detailed insight may even require numerical simulations.

\section*{Acknowledgments}
\addcontentsline{toc}{section}{Acknowledgements}
We thank Pengyu Lv for the inspiring discussions. We also thank the anonymous referees for the helpful suggestions to improve this paper. This work is part of an Industrial Partnership Programme (IPP) of the Netherlands Organization for Scientific Research (NWO). This research programme is co-financed by Canon Production Printing Holding B.V., University of Twente and Eindhoven University of Technology. DL gratefully acknowledges support by his ERCAdvanced Grant DDD (project number 740479).

\appendix
\section{Derivation of theoretical final volume}\label{appA}
The molar fraction of water in binary mixture is defined as the number of moles of water divided by the number of moles of both liquids. Here we neglect the tiny mole fraction ($<$ 0.1 \%) of silicone oil. The mole fraction of water is

\begin{equation}
x_w = \frac{m_w/M_w}{m_w/M_w + m_H/M_H} = \frac{1}{1 + \frac{m_H}{m_w}\frac{M_w}{M_H}}.
\label{xw}
\end{equation} 
where $m_w$ is the mass of water and $m_H$ is the mass of 1,2-hexanediol. Here we assume that the density of the mixture is linear with the solute concentration, which means that the total volume of the mixture is the sum of the individual liquid volumes~\citep{Battino1971}. This is a reasonable approximation for water/1,2-hexanediol mixture. Then the initial density is given by 

\begin{equation}
\rho_{i} = \frac{m_H + m_w}{\frac{m_H}{\rho_H} + \frac{m_w}{\rho_w}} = \frac{(m_H + m_w)\rho_H \rho_w}{m_H \rho_w + m_w \rho_H},
\label{rho_i}
\end{equation} 
where $\rho_i$ is the initial density of the mixture, $\rho_H$ and $\rho_w$ are the density of 1,2-hexanediol and water, respectively. We introduce the initial mass percentage $C$ of the solute, whose value can be between 0 and 1. Then we have $m_w = m_H\frac{1-C}{C}$ at the beginning before evaporation. By substituting it into Eq.~(\ref{rho_i}), the equation reduces to 

\begin{equation}
\rho_{i} = \frac{(m_H + m_H\frac{1-C}{C})\rho_H \rho_w}{m_H \rho_w + m_H\frac{1-C}{C} \rho_H}
= \frac{\rho_H \rho_w}{\rho_H + C(\rho_w - \rho_H)}.
\label{rho_i2}
\end{equation} 
The 1,2-hexanediol mass in the droplet is now given by $V_i \rho_i C$, which is constant during the drying process due to the non-volatility of 1,2-hexanediol. $V_i$ is defined as initial droplet volume. Therefore the amount of water that has evaporated is given by $(V_i - V)\rho_w$, and the total mass of water left in the droplet is the initial mass minus the evaporated mass, $V_i \rho_i (1 - C) - (V_i - V)\rho_w$. By substituting the water mass into Eq.~(\ref{xw}), the mole fraction of water in the droplet is expressed as

\begin{equation}
x_w = \frac{1}{1 + \frac{V_i\rho_i C}{V_i\rho_i(1 - C) - (V_i - V)\rho_w}\frac{M_w}{M_H}}.
\label{xw2}
\end{equation}  
From the theory, it is predicted that the droplet stops evaporating at the moment when the active mole fraction of water equals to relative humidity of surrounding air, $a_w = \psi_w x_w = RH$. Then we obtain the theoretical final mole fraction $x_w = RH/\psi_w$, where $RH$ and $\psi_w$ is the relative humidity and activity coefficient of water~\citep{marcolli2005}, respectively. By substituting $x_w$ into Eq.~(\ref{xw2}), we can analytically calculate the theoretical final volume $V_f$:

\begin{equation}
V_f = \left(\frac{M_w}{M_H}\frac{RH}{\psi_w - RH} + \frac{\rho_w}{\rho_H}\right)\left(\frac{1 - C}{C} + \frac{\rho_w}{\rho_H}\right)^{-1} V_0.
\label{vt}
\end{equation}


\end{document}